\newcommand{\CLASSINPUTtoptextmargin}{0.8in} 
\newcommand{\CLASSINPUTbottomtextmargin}{1in} 
\documentclass[conference]{IEEEtran}
\usepackage[mathcal]{eucal}
\usepackage{mathrsfs}

\usepackage{cite}

\ifCLASSINFOpdf
\usepackage[pdftex]{graphicx}
\else
  \usepackage[dvips]{graphicx}
\fi
\usepackage[cmex10]{amsmath}
\usepackage{amssymb}
\ifCLASSOPTIONcompsoc
  \usepackage[caption=false,font=normalsize,labelfont=sf,textfont=sf]{subfig}
\else
  \usepackage[caption=false,font=footnotesize]{subfig}
\fi

\usepackage{bm}
\usepackage{color}
\usepackage{amssymb}

\usepackage{epstopdf}
\newcommand{\argmin}{\mathop{\rm arg~min}\limits}

\begin{document}
%
\title{HoloCast: Graph Signal Processing for \\Graceful Point Cloud Delivery}

\author{\IEEEauthorblockN{Takuya Fujihashi\IEEEauthorrefmark{2}\IEEEauthorrefmark{1}\thanks{T. Fujihashi conducted this research while he was an intern at MERL.},
Toshiaki Koike-Akino\IEEEauthorrefmark{2},
Takashi Watanabe\IEEEauthorrefmark{3}, and
Philip V. Orlik\IEEEauthorrefmark{2}}
\IEEEauthorblockA{\IEEEauthorrefmark{2}Mitsubishi Electric Research Laboratories (MERL), Cambridge, MA 02139, USA}
\IEEEauthorblockA{\IEEEauthorrefmark{1}Graduate School of Science and Engineering, Ehime University, Ehime, Japan}
\IEEEauthorblockA{\IEEEauthorrefmark{3}Graduate School of Information and Science, Osaka University, Osaka, Japan}}


%


\maketitle

\begin{abstract}
In conventional point cloud delivery, a sender uses octree-based digital video compression to stream three-dimensional (3D) points and the corresponding color attributes over band-limited links, e.g., wireless channels, for 3D scene reconstructions. 
However, the digital-based delivery schemes have an issue called cliff effect, where the 3D reconstruction quality is a step function in terms of wireless channel quality.
We propose a novel scheme of point cloud delivery, called HoloCast, to gracefully improve the reconstruction quality with the improvement of wireless channel quality.
HoloCast regards the 3D points and color components as graph signals and directly transmits linear-transformed signals based on graph Fourier transform (GFT), without digital quantization and entropy coding operations. 
One of main contributions in HoloCast is that the use of GFT can deal with non-ordered and non-uniformly distributed multi-dimensional signals such as holographic data unlike conventional delivery schemes.
Performance results with point cloud data show that HoloCast yields better 3D reconstruction quality compared to digital-based methods in noisy wireless environment. 
\end{abstract}


%
\IEEEpeerreviewmaketitle

\section{Introduction}
\begin{figure}[t]
  \begin{minipage}{0.5\textwidth}
  \centering
   \subfloat[light detection and ranging (LIDAR) scenario~\cite{bib:LiDAR}]{\includegraphics[width=8.2cm]{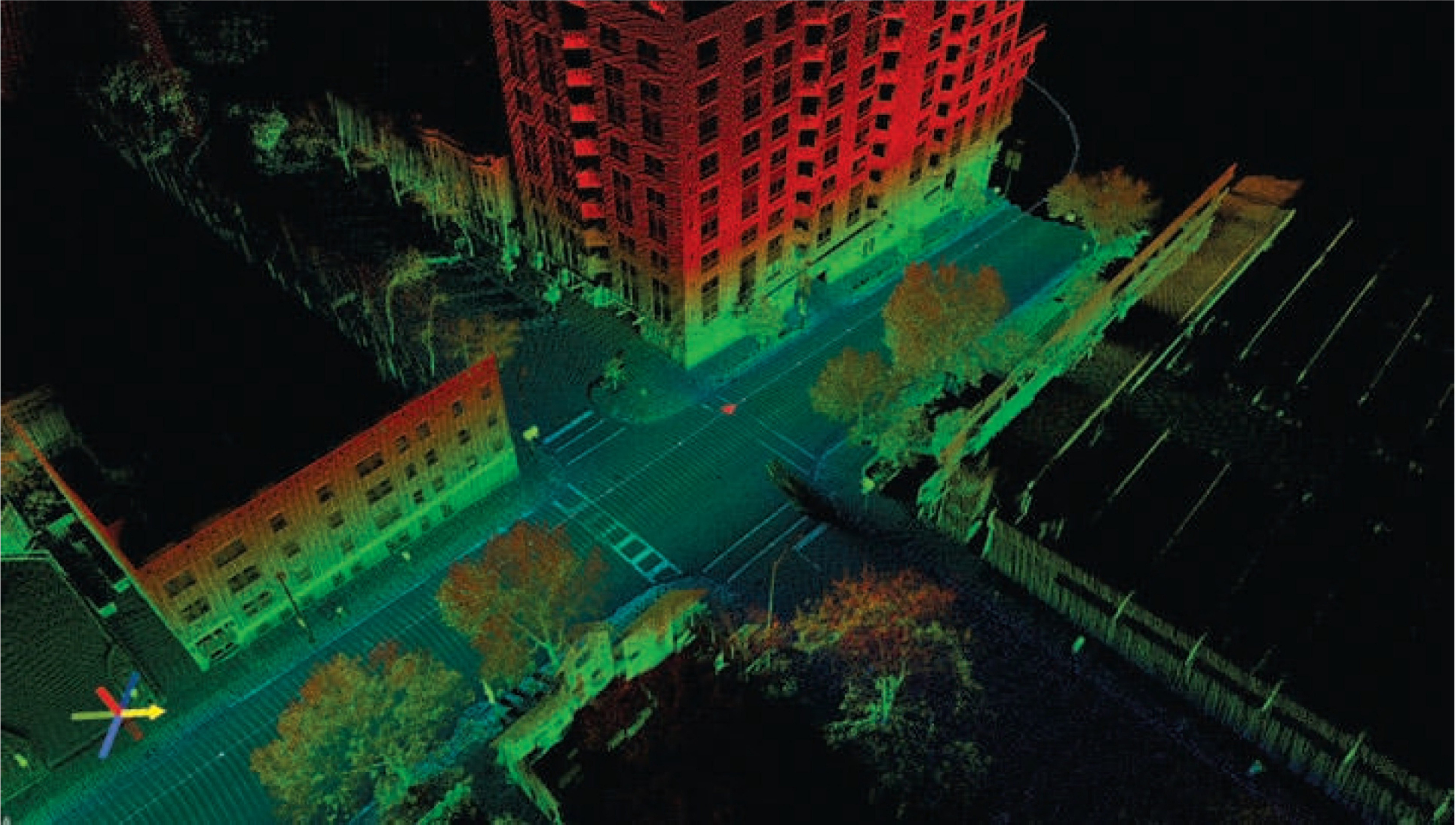}} \\
   \subfloat[AR/VR scenario~\cite{bib:AR}]{\includegraphics[width=8.2cm]{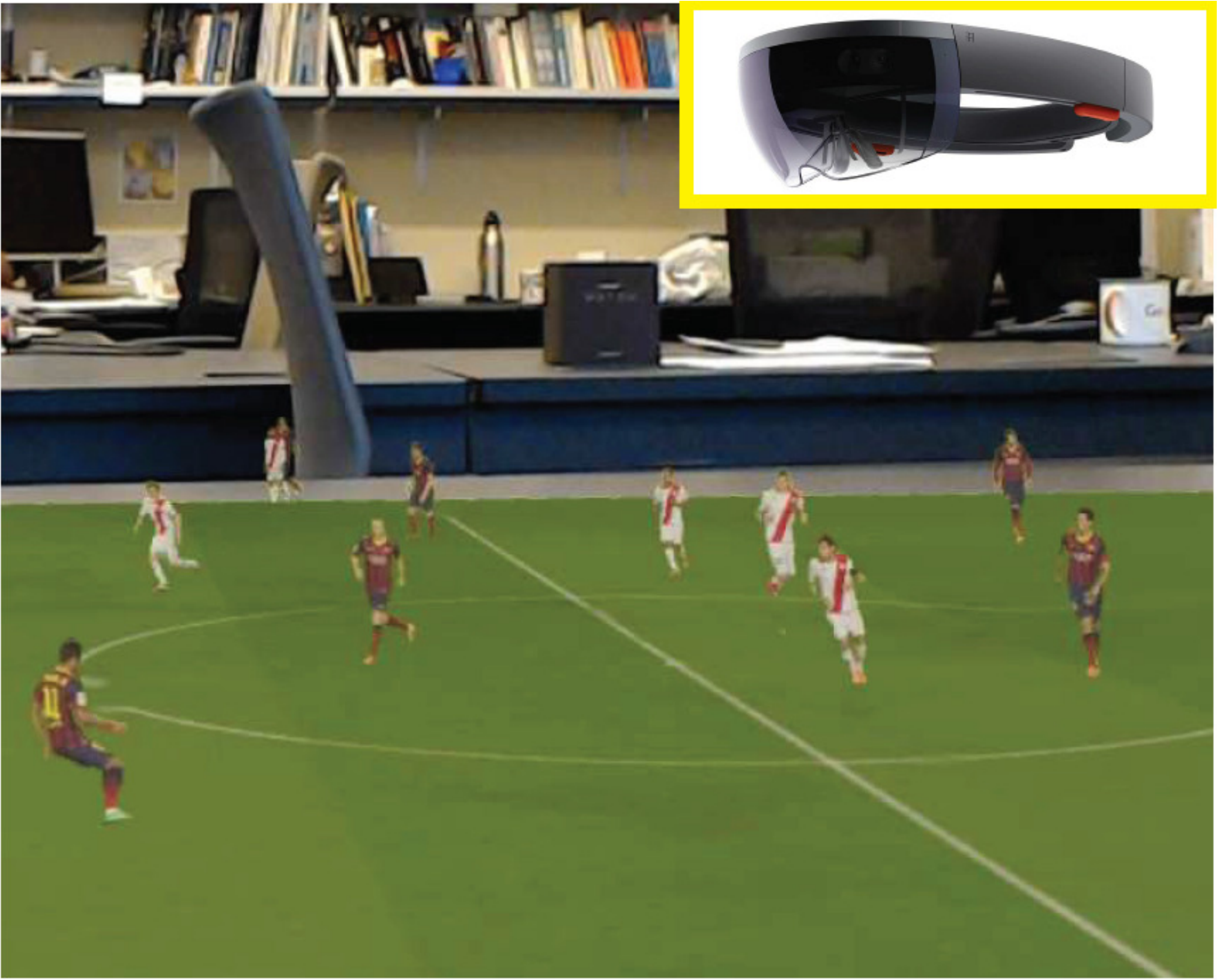}}
\end{minipage} \hfill
 \caption[]{Examples of holographic applications.} 
 \label{fig:pc}
\end{figure}

Holographic displays~\cite{bib:holodisplay,bib:holodisplay2} have emerged as attractive interface techniques for reconstructing three dimensional (3D) scenes that provide full parallax and depth information for human eyes.
3D holographic display can be widely used for many applications: entertainment, remote device operation, medical imaging, and simulated training as shown in Fig.~\ref{fig:pc}.
Point cloud~\cite{bib:mpeg} is one of data structures to reconstruct 3D scenes/objects on the holographic display~\cite{bib:hologram}. 
Point cloud is a set of 3D points, and each point is defined by 3D coordinates, i.e., (X, Y, Z) and color attributes, i.e., (R, G, B) or (Y, U, V).

In contrast to conventional two dimensional (2D) images, 3D points in point cloud data are not ordered and are non-uniformly distributed in space.
One of major issues in point cloud delivery is how to compress and send such numerous and irregular structure of 3D points while maintaining high 3D reconstruction quality on displays. 
For example, when the number of 3D points is $800{,}000$, the amount of traffic without any compression is approximately $38$~Mbits~\cite{bib:datasize}. 
Large traffic causes low reconstruction quality in point cloud delivery over limited data rate links, especially, wireless communications. 

\begin{figure}[t]
 \begin{center}
  \includegraphics[width=\hsize]{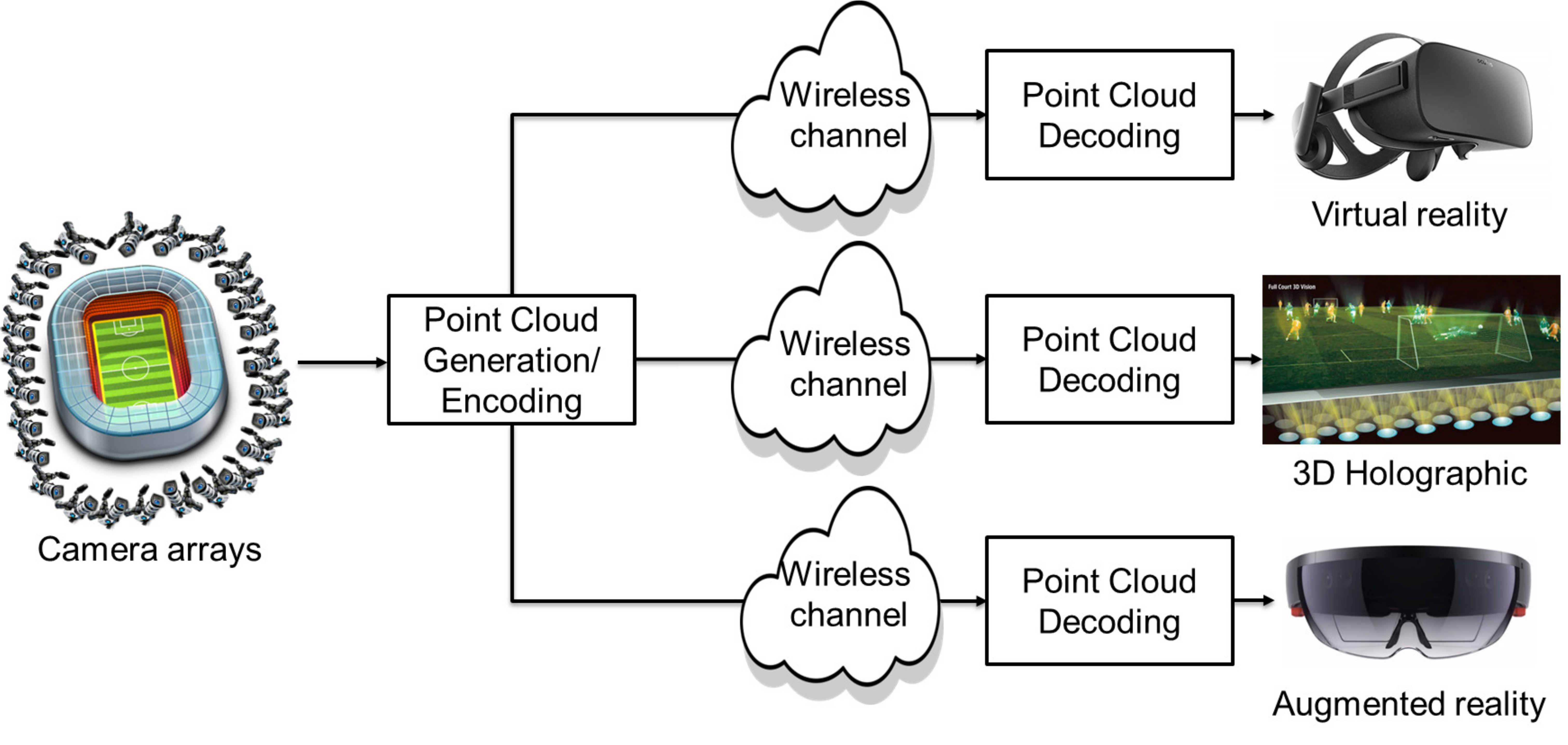}
  \caption{Wireless point cloud delivery for immersive video applications.}
  \label{fig:vrar}
 \end{center}
\end{figure}

For point cloud compression over wireless links, conventional encoders, such as popular Point Cloud Library (PCL)~\cite{bib:PCL, bib:PCL2}, use octree decomposition, prediction, quantization, and entropy coding. 
Specifically, a sender first decomposes point cloud into multiple 3D point sets~\cite{bib:octree} and takes quantization and entropy coding for each point set to generate the compressed bitstream for transmissions. 
Here, the compression rate of the bitstream is adaptively selected according to the wireless channel quality.
After the compression, the transmission part uses a channel coding and digital modulation scheme to
reliably transmit the compressed bitstream over wireless channels.
High-quality transmissions of point clouds over wireless links can realize immersive video applications such as virtual reality and augmented reality on wireless devices as shown in Fig.~\ref{fig:vrar}.

However, the conventional schemes of point cloud delivery suffer from the following problems due to the
wireless channel unreliability.  First, the encoded bitstream is highly
vulnerable for bit errors~\cite{bib:survey}.  When the channel signal-to-noise ratio (SNR)
falls under a certain threshold, possible few bit errors occurred in the bitstream during communications can cause a synchronization problem in point cloud decoding.
As a result, the display does not reconstruct 3D scenes, and thus the reconstruction quality degrades significantly. This phenomenon is called cliff effect~\cite{bib:fuji_GMRF}. 
Second, the reconstruction quality does not improve 
even when the wireless channel quality is improved unless an adaptive rate control of source and channel coding is performed in real-time according to the rapid fading channels. This is called leveling effect.  Finally,
quantization is a lossy process and its distortion cannot be recovered
at the receiver.

As mentioned above, conventional point cloud transmissions have two challenging issues over wireless links: 1) cliff effect and 2) leveling effect.
To overcome these issues, we propose a new point cloud transmission scheme to reconstruct 3D scenes in high quality on holographic displays.
The key idea of this scheme is skipping nonlinear operations, i.e., quantization and entropy coding, in point cloud coding.
For high-quality delivery, this study considers 3D points as vertices in a graph $\mathcal{G}$, with edges between nearby vertices to deal with irregular structure motivated by~\cite{bib:graph_PCC,bib:graph_PCC2}. 
Each point $\boldsymbol{p}$ has attributes of 3D coordinates and color components, and those attributes are regarded as signals residing on the vertices of the graph.
The proposed scheme takes graph Fourier transform (GFT)~\cite{bib:GSP,bib:GSIP} for each attribute in graph signals to compact the signal power, whose output is then scaled and directly mapped to transmission signals without relying on digital modulation schemes.
The advantage of this modification lies in a fact that the point distortion due to communication noise is proportional to the magnitude of the noise, resulting into graceful reconstruction quality according to the wireless channel quality, without any cliff effect and leveling effect.
We demonstrate that the proposed point cloud delivery scheme achieves graceful reconstruction quality with the improvement of wireless channel quality and better reconstruction performance compared to the conventional digital-based schemes.
For example, HoloCast achieves $39.45$~dB and $32.48$~dB improvement in the attributes of the 3D coordinates and color components, respectively, in terms of mean squared error (MSE) compared with the digital-based delivery schemes.

\paragraph*{Related Works and Our Contributions}
Soft video delivery schemes have been recently proposed for multi-dimensional ordered video signals in~\cite{bib:SoftCast_second,bib:foveacast, bib:swift, bib:FreeCast}.
For example, SoftCast~\cite{bib:SoftCast_second} was designed for 3D ordered video signals to realize graceful video delivery. They skip quantization and entropy coding, and uses 3D discrete-cosine transform (DCT) and analog modulation, which maps DCT coefficients directly to transmission signals, to ensure that the received video quality is proportional to wireless channel quality. FoveaCast~\cite{bib:foveacast} considers the foveation characteristic of human vision into soft video delivery of 2D ordered video signals to achieve higher visual perceptual quality.
FreeCast~\cite{bib:FreeCast} extended the soft video delivery towards 5D ordered multi-view video plus depth (MVD) signals. They use 5D-DCT for decorrelation and directly send the coefficients to realize graceful quality improvement with the improvement of wireless channel quality.

Our study realizes soft coding and decoding for point cloud delivery. Although existing schemes of soft video delivery deal with ordered and uniformly distributed video signals, point cloud delivery needs to handle non-ordered and non-uniformly distributed points in coding and decoding. To this end, the proposed scheme has the following major contributions: 
 \begin{itemize}
 \item We regard point clouds as graph signals with the attributes of 3D coordinates and color components to deal with irregular structure of holographic data formats.

 \item We introduce GFT and analog modulation for graph signals to exploit correlations in graph-domain for performance improvement.
            
 \item We discuss an impact of graph Laplacian variants and adjacency hyperparameters on 3D scene reconstruction quality. 

  \item We demonstrate that GFT-based HoloCast achieves graceful 3D reconstruction quality with a significant performance improvement over digital-based point cloud delivery.
\end{itemize}

\begin{figure*}[t]
 \begin{center}
  \includegraphics[width=\hsize]{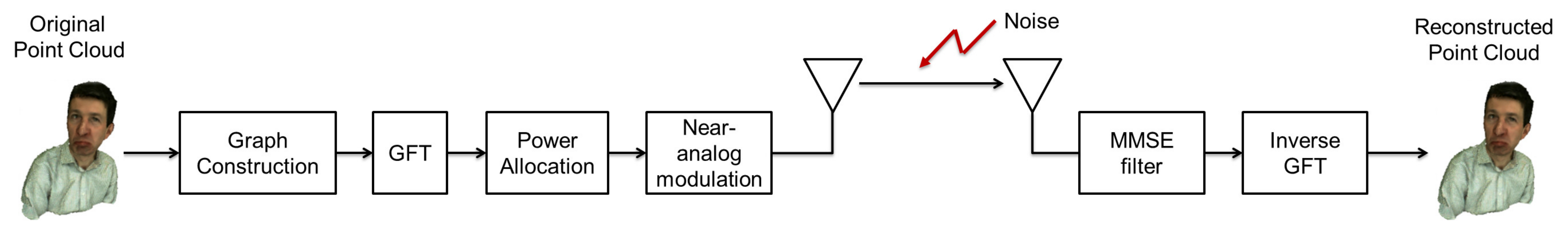}
  \caption{Overview of HoloCast for graceful point cloud delivery.}
  \label{fig:scheme}
 \end{center}
\end{figure*}

\section{HoloCast: Graceful Point Cloud Delivery}

The objectives of our study are 1) to prevent cliff effect in 3D scene reconstruction and 2) to gracefully improve reconstruction quality 
with the improvement of wireless channel quality.

Fig.~\ref{fig:scheme} shows the overview of proposed HoloCast.
The encoder first performs GFT for 3D points and the corresponding colors, i.e., graph signals. 
The GFT coefficients are then scaled and analog-modulated according to the signal power information for wireless transmissions.  
Next, the encoder sends the analog modulated symbols to the receiver over a wireless channel, which is often impaired with additive white Gaussian noise (AWGN) and time-varying fading.
At the receiver side, the decoder uses minimum mean-square error (MMSE) filter to obtain the transmitted GFT coefficients.
The decoder finally takes inverse GFT to reconstruct 3D coordinates and color components for display.

\subsection{Encoder}
We first represent 3D points and color components using a weighted and undirected graph $\mathcal{G} = (\boldsymbol{V}, \mathcal{E}, \boldsymbol{W})$ where $\boldsymbol{V}$ and $\mathcal{E}$ are the vertex and edge sets of $\mathcal{G}$, respectively.  $\boldsymbol{W}$ is an adjacency matrix having positive edge weights and the $(i,j)$th entry $\boldsymbol{W}_{i, j}$ represents the weight of an edge connecting vertices $i$ and $j$. 
For the graph $\mathcal{G} = (\boldsymbol{V}, \mathcal{E}, \boldsymbol{W})$, we consider the attributes of the point cloud, i.e., the 3D coordinates $\bm{p} = [\mathsf{x}, \mathsf{y}, \mathsf{z}]^\mathrm{T} \in \mathbb{R}^{3 \times N}$ and the color components $\bm{c} = [y, u, v]^\mathrm{T} \in \mathbb{R}^{3 \times N}$ as signals that reside on the vertices in the graph ($N$ is the number of vertices). From the attributes, each weight $\boldsymbol{W}_{i, j}$ can be calculated, e.g., by the Gaussian kernel as follows:
\begin{equation}
\boldsymbol{W}_{i, j} = \exp \left (-\frac{||\bm{p}_i - \bm{p}_j||^2_2}{\kappa} \right ),
 \label{eq:W}
\end{equation}
where $\bm{p}_i$ represents the 3D coordinates of point $i$ and $\kappa$ is a hyperparameter. In HoloCast, we use either the sample variance or the standard deviation of distances across all the points for the hyperparameter $\kappa$. 
A sender then transforms the graph signals into spectral representation using GFT. 
The GFT is defined through the graph Laplacian operator $\boldsymbol{L}$ using edge weight matrix $\boldsymbol{W}$ and degree matrix $\boldsymbol{D}$, where $\boldsymbol{D}$ is the diagonal degree matrix whose $i$th diagonal element is equal to the sum of the weights of all the edges incident to vertex $i$. Specifically, the diagonal matrix is represented as: 
\begin{equation}
\boldsymbol{D}_{i,j} = 
\begin{cases}
\sum_{n=1}^{N} \boldsymbol{W}_{i,n}, & \text{if} \; i = j, \\
0, &\text{otherwise}.
\end{cases}
\end{equation}
Based on the degree matrix, we can calculate some variants of the graph Laplacian matrix~\cite{bib:Laplacian}:
\begin{align}
\boldsymbol{L} &= \boldsymbol{D} - \boldsymbol{W}, \\
\boldsymbol{L} &= \boldsymbol{I} - \boldsymbol{D}^{-1/2} \boldsymbol{W} \boldsymbol{D}^{-1/2}, \\
\boldsymbol{L} &= \boldsymbol{D}^{-1} \boldsymbol{W}, \\
\boldsymbol{L} &= \boldsymbol{I}- \boldsymbol{D}^{-1} \boldsymbol{W},
\end{align}
where $\boldsymbol{I}$ denotes an identity matrix of proper dimension.
We refer to each graph Laplacian matrix as regular, normalized, transition, and random-walk Laplacian, respectively.  
We will discuss the impact of those graph Laplacian matrices on the delivery quality in Section~\ref{sec:laplacian}.
In general, the graph Laplacian is a real symmetric matrix that has a complete set of orthonormal eigenvectors with corresponding nonnegative eigenvalues. To obtain the eigenvectors and eigenvalues, the eigen decomposition of the Laplacian matrix is performed as: 
\begin{equation}
\boldsymbol{L} = \boldsymbol{\varPhi} \boldsymbol{\varDelta} \boldsymbol{\varPhi}^{-1},
\end{equation}
where $\boldsymbol{\varPhi}$ is the eigenvectors matrix and $\boldsymbol{\varDelta}$ is a diagonal matrix containing the eigenvalues.\footnote{For non-diagonalizable graph Laplacian matrix, the singular value decomposition (SVD) is instead used to express as $\boldsymbol{L} = \boldsymbol{\varPsi} \boldsymbol{\varDelta} \boldsymbol{\varPhi}^{-1}$ where $\boldsymbol{\varPsi}$, $\boldsymbol{\varDelta}$ and $\boldsymbol{\varPhi}$ denote left singular vectors matrix, diagonal matrix containing singular values, and right singular vectors matrix, respectively. In this case, we use the right singular vectors of $\boldsymbol{\varPhi}$ as the graph-based transform basis matrix $\boldsymbol{\varPhi}$.}
The multiplicity of the smaller eigenvalue indicates the number of connected components of the graph. The GFT coefficients of each attribute of graph signals $\bm{f} \in R^N$ are obtained by multiplying the graph-based transform basis matrix by the corresponding attribute vector as follows:
\begin{equation}
\bm{s} = \bm{f} \boldsymbol{\varPhi},
\end{equation}
where $\bm{s}$ is a vector of GFT coefficients corresponding to the graph signals of $\bm{f}$.
After power allocation for each GFT coefficient, the GFT coefficients are mapped to I (in-phase) and Q (quadrature-phase) components for analog wireless transmissions.

Let $x_{i}$ denote the $i$th analog-modulated symbol, which is the $i$th GFT coefficient $s_i$ of all the attributes scaled by a factor of $g_{i}$ for noise reduction as follows:
\begin{equation}
x_{i} = g_{i} \cdot s_{i}.
\end{equation}
The optimal scale factor $g_{i}$ is obtained by minimizing the MSE under the power constraint with a total power budget of $P$ as follows:
\begin{align}
\mathop{\min}_{\{g_i\}} &\quad \mathsf{MSE} =
 \mathbb{E} \left [ \left (s_{i} - \hat{s}_{i}\right)^2\right] = \sum_{i}^{N} \frac{\sigma^2 {\lambda}_{i}}{g_{i}^2{\lambda}_{i} + \sigma^2},\\
\mathrm{s.t.} &\quad \frac{1}{N}\sum_{i}^{N}  g_{i}^2{\lambda}_{i} = P,
\end{align}
where $\mathbb{E}[\cdot]$ denotes expectation, $\hat{s}_{i}$ is a receiver estimate of the transmitted GFT coefficient, ${\lambda}_{i}$ is the power of the $i$th GFT coefficient, $N$ is the number of GFT coefficients, and $\sigma^2$ is a receiver noise variance.
As shown in~\cite{bib:SoftCast_second}, the near-optimal solution is expressed as
\begin{equation}
\label{water}
g_{i} = {\lambda}_{i}^{-1/4} \sqrt{\frac{NP}{\sum_j\sqrt{\lambda}_{j}}}.
\end{equation}

\subsection{Decoder}

Over the wireless links, the receiver obtains the received symbol, which is modeled as follows:
\begin{equation}
y_{i} = x_{i} + n_{i},
\end{equation}
where $y_{i}$ is the $i$th received symbol and $n_{i}$ is an effective AWGN with a variance of $\sigma^2$ (which is already normalized by wireless channel strength in the presence of fading attenuation).
The GFT coefficients are extracted from I and Q components via an MMSE filter~\cite{bib:SoftCast_second}:
\begin{equation}
\hat{s}_{i} = \frac{g_{i} {\lambda}_{i}}{g_{i}^2 {\lambda}_{i} + \sigma^2} \cdot y_{i}.
 \label{eq:mmse}
\end{equation}
The decoder then reconstructs corresponding graph signals $\hat{\bm{f}}$, i.e., attributes of 3D coordinates and color components, by taking the inverse GFT for the filtered GFT coefficients in each attribute $\hat{\bm{s}}$ as follows:
\begin{equation}
\hat{\bm{f}} = \hat{\bm{s}} \, \boldsymbol{\varPhi}^{-1}.
\end{equation}

\subsection{Analog Compression for Limited Bandwidth}
The previous designs assume that the sender has enough bandwidth to transmit all the coefficients in the spectral domain over the wireless medium. 
If the available bandwidth and/or time resources are restricted for wireless channel use, it has to selectively transmit the coefficients to fit the available bandwidth.
For such cases, our scheme sorts the coefficients in descending order of the power and picks higher-power coefficients to fill the bandwidth.
When the sender discards a coefficient, the receiver regards the discarded coefficient as zero. As a result, a sort of data compression can be accomplished even for analog-based video delivery. 
Even when some coefficients are discarded to reduce the amount of data, the receiver can still achieve a graceful video quality until reaching the distortion limit due to the compression.

\section{Performance Evaluation}
\subsection{Simulation Settings}
\label{sec:settings}
\noindent \textbf{Performance Metric:} We evaluate the reconstruction quality of point cloud delivery in terms of the symmetric MSE based on~\cite{bib:symMSE} in each attribute of 3D coordinates $\bm{p}$ and color components $\bm{c}$. The symmetric MSE of the 3D coordinates, $\mathsf{sMSE}_\mathsf{xyz}$, can be obtained as follows:
\begin{equation}
\mathsf{sMSE}_\mathsf{xyz} = \frac{1}{2} \bigl( d(\bm{p}_\mathrm{org} \to \bm{p}_\mathrm{dec}) +  d(\bm{p}_\mathrm{dec} \to \bm{p}_\mathrm{org}) \bigr),
\end{equation}
where $\bm{p}_\mathrm{org}$ is the original 3D coordinates and $\bm{p}_\mathrm{dec}$ is the decoded 3D coordinates. Here, each way of the asymmetric MSE in the 3D coordinates are defined as follows:
\begin{eqnarray}
d(\bm{p}_\mathrm{org} \to \bm{p}_\mathrm{dec}) = \frac{1}{N} \sum_{\bm{p} \in \bm{p}_\mathrm{org}} \left (\min_{\bm{p}' \in \bm{p}_\mathrm{dec}}\bigl\| \bm{p} - \bm{p}'\bigr\|_2^2 \right ), \\
d(\bm{p}_\mathrm{dec} \to \bm{p}_\mathrm{org}) = \frac{1}{N} \sum_{\bm{p} \in \bm{p}_\mathrm{dec}} \left (\min_{\bm{p}' \in \bm{p}_\mathrm{org}} \bigl\| \bm{p} - \bm{p}' \bigr\|_2^2 \right ).
\end{eqnarray}

The symmetric MSE of the color components, $\mathsf{sMSE}_\mathsf{yuv}$, is derived analogously as follows:
\begin{equation}
\mathsf{sMSE}_\mathsf{yuv} = \frac{1}{2} \bigl( d(\bm{c}_\mathrm{org} \to \bm{c}_\mathrm{dec}) +  d(\bm{c}_\mathrm{dec} \to \bm{c}_\mathrm{org}) \bigr),
\end{equation}
where $\bm{c}_\mathrm{org}$ and $\bm{c}_\mathrm{dec}$ are the original and decoded color components, respectively.
In this case, the asymmetric MSE of the color component is defined as follows:
\begin{multline}
d(\bm{c}_\mathrm{org} \to \bm{c}_\mathrm{dec}) = \frac{1}{N} \sum_{\bm{c} \in \bm{c}_\mathrm{org}} \left ( \bigl\| \bm{c} - \bm{c}_\mathrm{dec}(\bm{p}_\mathrm{min}') \bigr\|_2^2 \right ), \\
\bm{p}_\mathrm{min}' = \argmin_{\bm{p}' \in \bm{p}_\mathrm{dec}} \bigl\| \bm{p}_\mathrm{org} - \bm{p}' \bigr\|_2^2, 
\end{multline}
\begin{multline}
d(\bm{c}_\mathrm{dec} \to \bm{c}_\mathrm{org}) = \frac{1}{N} \sum_{\bm{c} \in \bm{c}_\mathrm{dec}} \left (\bigl\| \bm{c} - \bm{c}_\mathrm{org}(\bm{p}_\mathrm{min}'') \bigr\|_2^2 \right ), \\
\bm{p}_\mathrm{min}'' = \argmin_{\bm{p}'' \in \bm{p}_\mathrm{org}} \bigl\| \bm{p}_\mathrm{dec} - \bm{p}'' \bigr\|_2^2, 
\end{multline}
where $\bm{c}_\mathrm{dec/org}(\bm{p})$ represents the color components of the corresponding 3D coordinates $\bm{p}$.

\noindent \textbf{Point Cloud Dataset:} We use the reference point clouds, namely,
{\it pencil\_\_10\_0}, {\it pencil\_\_9\_0}, and {\it office1}, whose number of points $N$ is $2{,}731$, $6{,}712$, and $307{,}200$, respectively. 
We first focus on {\it pencil\_\_10\_0} to compare between HoloCast and digital-based delivery. The performance at a different number of 3D points will then be evaluated with {\it pencil\_\_9\_0}. In addition, we demonstrate the visual quality for the point cloud data of {\it office1} in Section~\ref{sec:visual}.

\noindent \textbf{Wireless Settings:} The received symbols are impaired by an AWGN channel. 
For digital-based schemes, we use a rate-$1/2$ convolutional codes with a constraint length of $10$. The digital modulation formats are either quadrature phase-shift keying (QPSK), $16$-ary quadrature-amplitude modulation (16QAM), or 256-ary QAM (256QAM).

\noindent \textbf{Digital Point Cloud Coder:} We compare HoloCast with the conventional digital-based delivery, which is based on point cloud digital compression used in PCL~\cite{bib:PCL}. We consider two default profiles for compression: LOW\_RES\_OFFLINE\_COMPRESSION\_WITH\_COLOR and MED\_RES\_OFFLINE\_COMPRESSION\_WITH\_COLOR.
Note that this digital-based PCL point cloud delivery does not exploit GFT, while there exist recent work of GFT-based digital compression schemes to improve the efficiency, e.g., \cite{bib:graph_PCC,bib:graph_PCC2}. 
Because the primary objective of our paper is a preliminary demonstration of a new soft delivery technique for point cloud data, 
we focus on widely used PCL-based digital delivery for benchmark performance comparisons in this paper.
Nevertheless, we plan to compare our HoloCast with GFT-based digital compression methods in near future.

\begin{figure}[t]
  \begin{minipage}{0.5\textwidth}
  \centering
   \subfloat[3D coordinates $\boldsymbol{p}$]{\includegraphics[width=8.8cm]{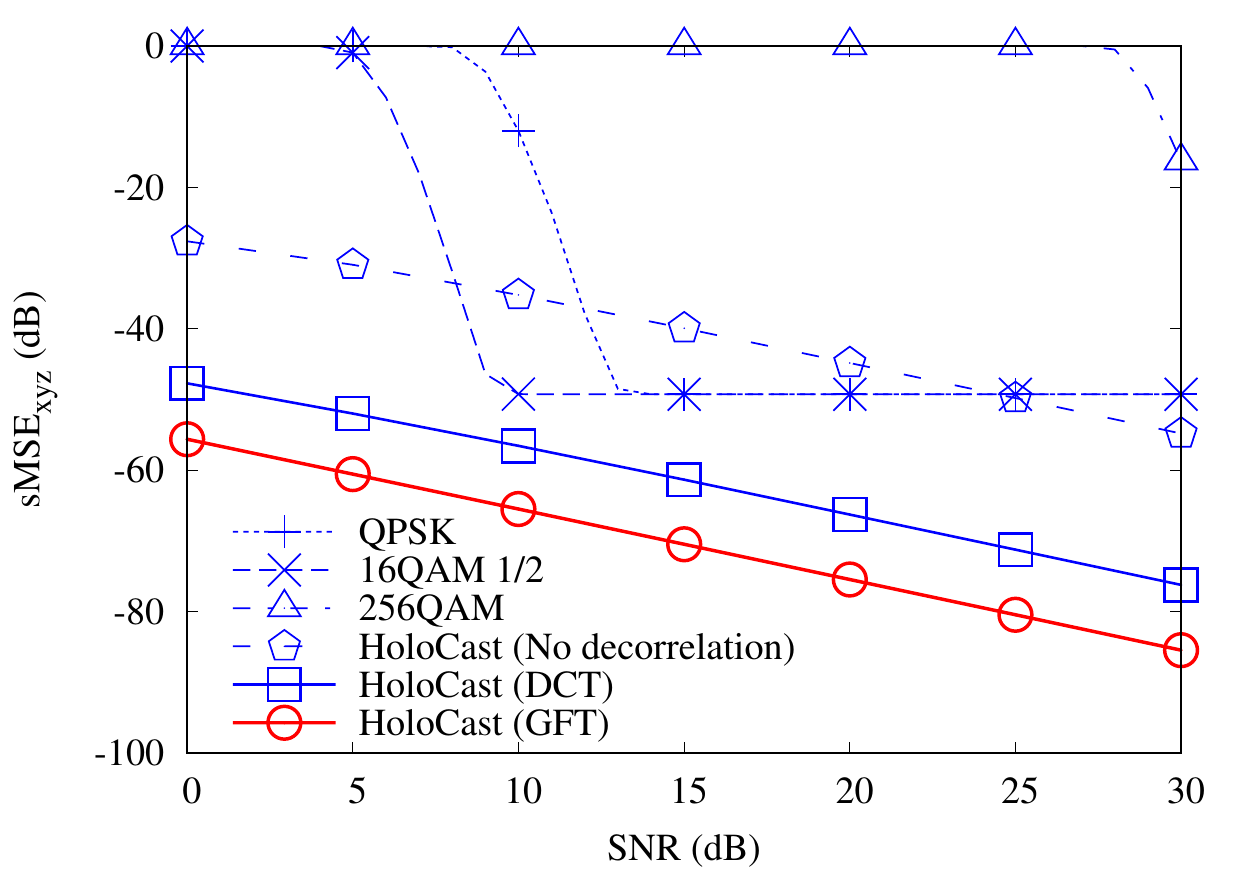}} \\
   \subfloat[Color components $\boldsymbol{c}$]{\includegraphics[width=8.8cm]{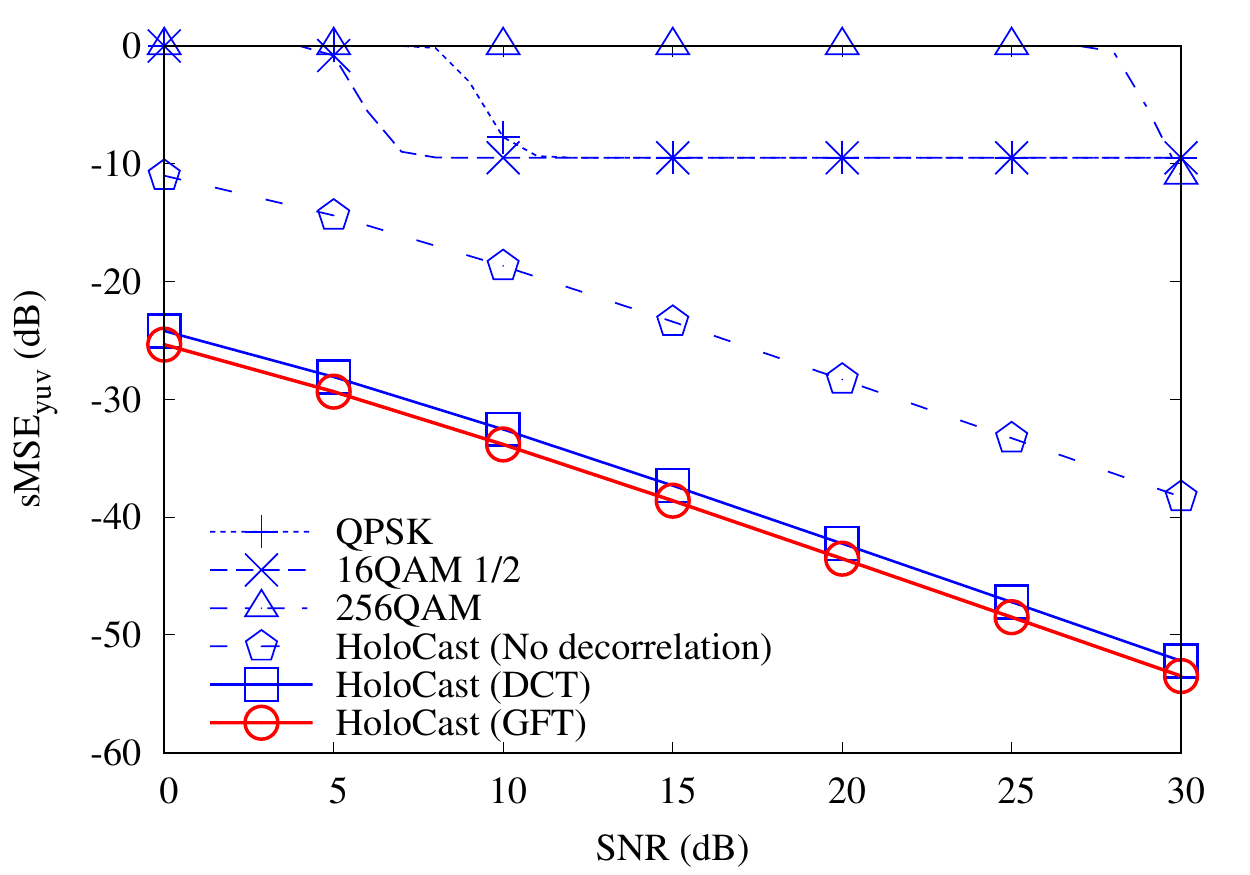}}
\end{minipage} \hfill
  \caption{MSE of 3D coordinates and color attributes in digital-based schemes and HoloCast for \textit{pencil\_\_10\_0} ($N=2731$).}
  \label{fig:snrvsmse}
\end{figure}

\subsection{HoloCast vs. Digital-based Schemes}
We first evaluate the quality of HoloCast and conventional digital-based schemes.
Fig.~\ref{fig:snrvsmse}~(a) shows the MSE of 3D coordinate attributes $\boldsymbol{p}$ in the digital-based scheme and HoloCast as a function of wireless channel SNRs. Here, HoloCast uses the sample variance of point distances as the hyperparameter $\kappa$ and the random-walk matrix for the graph Laplacian $\boldsymbol{L}$. In addition, we consider additional two HoloCast schemes to demonstrate an impact of GFT on quality improvement: DCT-based decorrelation and no decorrelation.
From evaluation results in Fig.~\ref{fig:snrvsmse}~(a), we can see the following observations:
\begin{itemize}
\item HoloCast gracefully improves the reconstruction quality of 3D coordinate attributes with the improvement of wireless channel quality.
\item Digital-based schemes suffer from cliff effect at low channel SNR regimes because bit errors cause synthesis errors of entropy decoding and leveling effect at high channel SNR regimes due to quantization errors.
\item GFT-based HoloCast can achieve better MSE compared with DCT-based HoloCast and HoloCast w/o decorrelation. GFT can utilize correlations of non-ordered and non-uniformly distributed 3D points by treating the 3D point data as graph signals.
\end{itemize}
For example, HoloCast achieves $30.1$~dB and $8.9$~dB improvement compared with HoloCast without decorrelation and DCT-based HoloCast on average across the channel SNRs between 0~dB and 30~dB, respectively.

Fig.~\ref{fig:snrvsmse}~(b) also shows the MSE of color component attributes in the digital-based scheme and HoloCast as a function of wireless channel SNRs. Even in the attributes of the color components, HoloCast realizes graceful quality improvement with the improvement of wireless channel quality. In digital-based schemes, they have low reconstruction quality even in high channel SNR regimes. It suggests that GFT-based decorrelation has a great advantage to represent point clouds with higher reconstruction quality.  

For further quality improvement in color components, we can consider the bilateral Gaussian kernel~\cite{bib:bilateral} in each weight $\boldsymbol{W}_{i, j}$ to decorrelate color components more efficiently. Specifically, Eq.~(\ref{eq:W}) will be modified as follows: 
\begin{equation}
\boldsymbol{W}_{i, j} = \exp \left (- \left (\frac{\|\bm{p}_i - \bm{p}_j\|^2_2}{\kappa_{p}} + \frac{\|\bm{c}_i - \bm{c}_j\|^2_2}{\kappa_{c}}\right ) \right ),
 \label{eq:bi}
\end{equation}
where  $\kappa_{p}$ and $\kappa_{c}$ are hyperparameters for 3D coordinates and color components, respectively.
Our evaluation in the quality of color components verified that the use of bilateral kernel in (\ref{eq:bi}) instead of (\ref{eq:W}) can offer additional $6.64$~dB gain on average across the channel SNRs between 0~dB and 30~dB.

\begin{figure}[t]
  \begin{minipage}{0.5\textwidth}
  \centering
   \subfloat[3D coordinates $\boldsymbol{p}$]{\includegraphics[width=8.8cm]{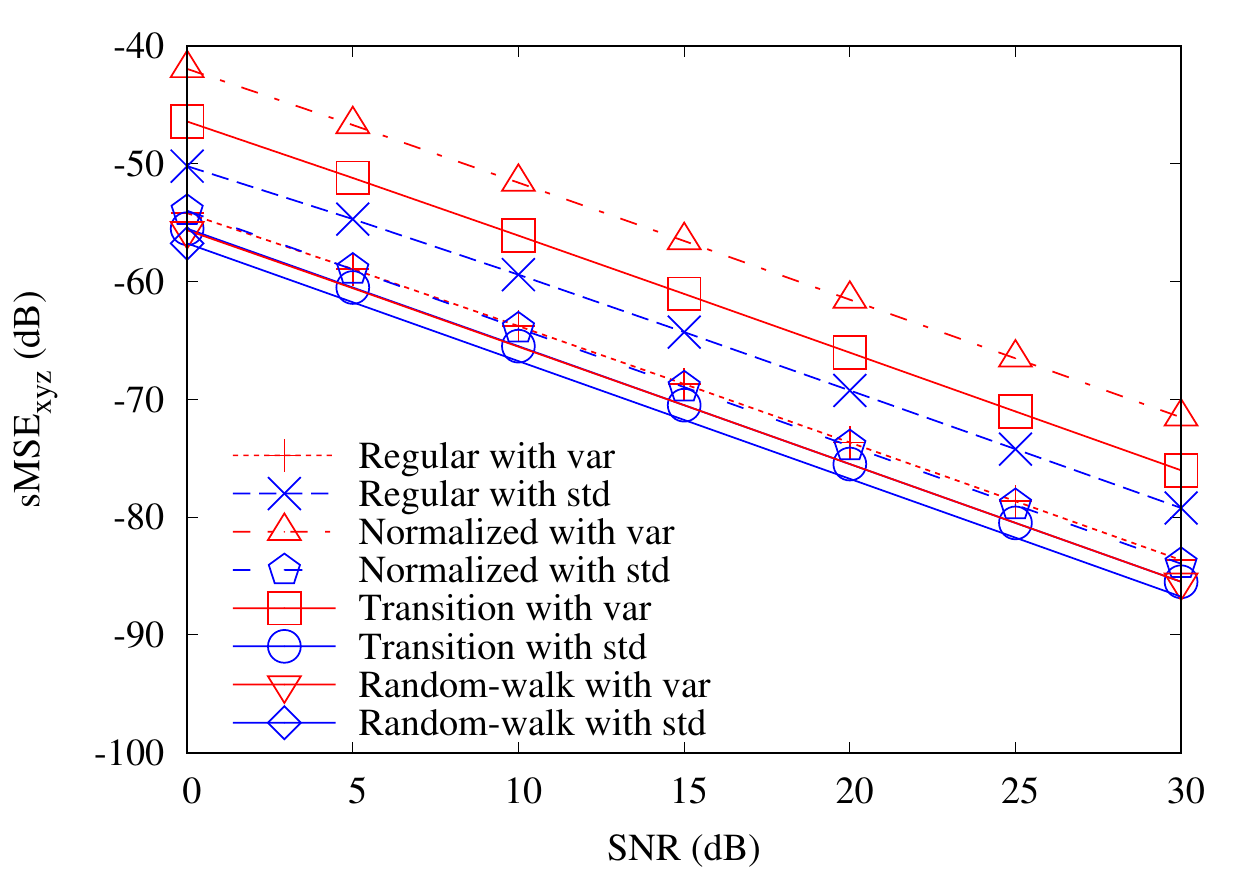}} \\
   \subfloat[Color components $\boldsymbol{c}$]{\includegraphics[width=8.8cm]{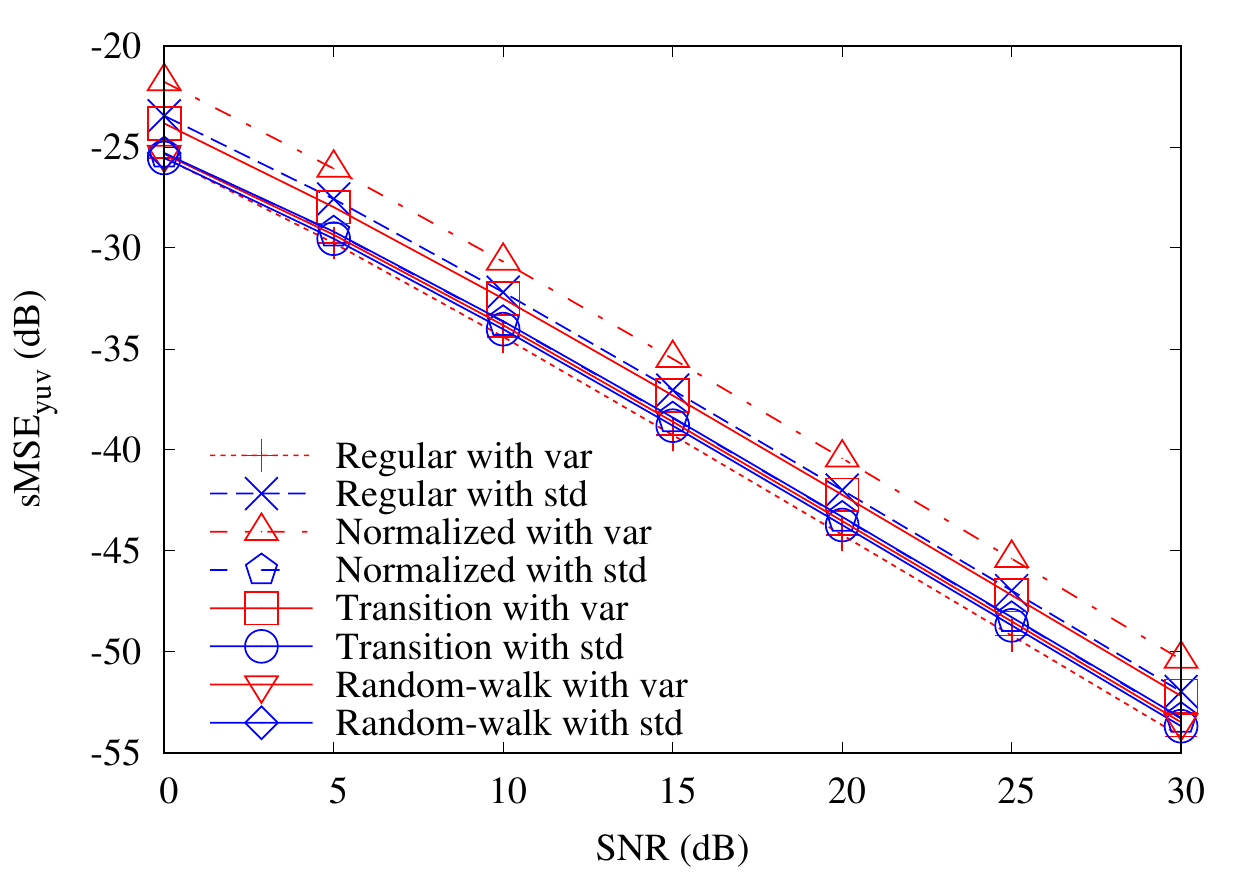}}
\end{minipage} \hfill
  \caption{MSE of 3D coordinates and color attribute in HoloCast with different graph Laplacian matrix and hyperparameters for \textit{pencil\_\_10\_0} ($N=2731$).}
  \label{fig:hyper}
\end{figure}

\subsection{Impacts of Graph Laplacian Matrix and Adjacency Hyperparameters}
\label{sec:laplacian}
In the previous section, we evaluated the performance of HoloCast using the random-walk graph Laplacian matrix $\boldsymbol{L}$ and variance-based hyperparameter $\kappa$.
In HoloCast, different types of graph Laplacian matrix can be used to encode/decode graph signals. In addition, the weight matrix $\boldsymbol{W}$ under consideration in Eq.~(\ref{eq:W}) highly depends on the value of Gaussian kernel hyperparameter $\kappa$. For the calculation of $\kappa$, the sample variance (var) or the standard deviation (std) of point distances is often used.   
In this section, we discuss the effects of graph Laplacian matrix and hyperparameter on the reconstruction quality in detail.

\begin{figure}[t]
  \begin{minipage}{0.5\textwidth}
  \centering
   \subfloat[3D coordinates $\boldsymbol{p}$]{\includegraphics[width=8.8cm]{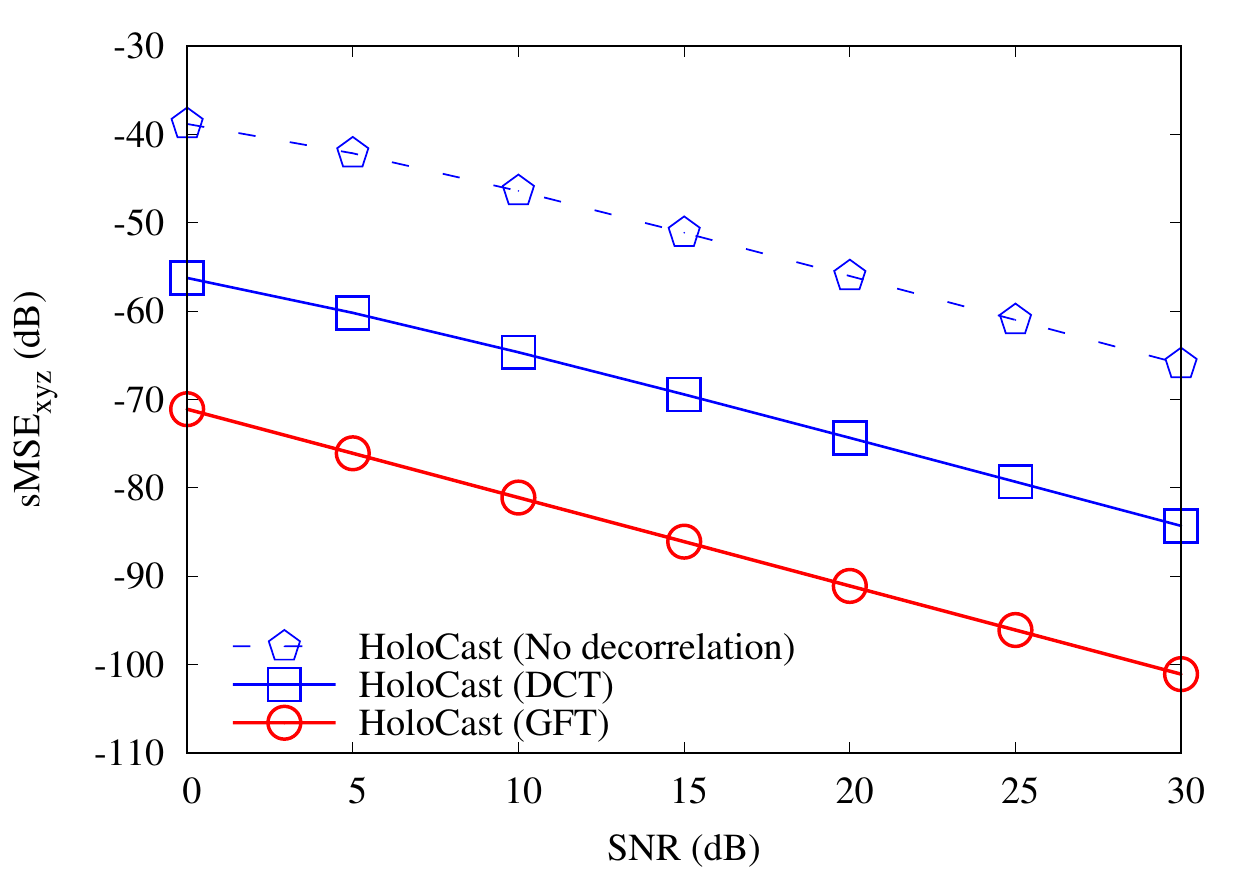}} \\
   \subfloat[Color components $\boldsymbol{c}$]{\includegraphics[width=8.8cm]{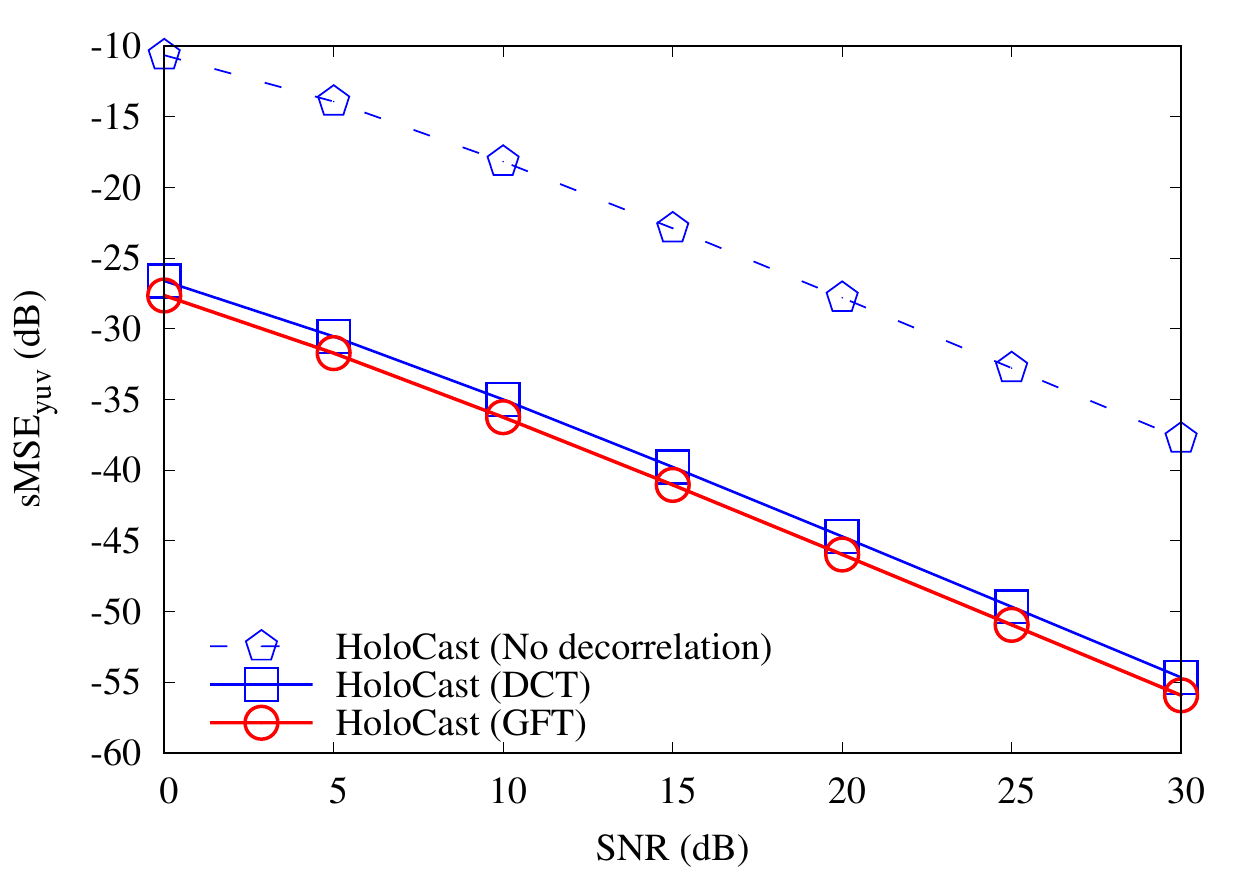}}
\end{minipage} \hfill
  \caption{MSE of 3D coordinates and color attribute in HoloCast with different decorrelation transform for {\it pencil\_\_9\_0} ($N=6712$).}
  \label{fig:snrvsmse_9_0}
\end{figure}

Figs.~\ref{fig:hyper}~(a) and (b) show the MSE of 3D coordinate and color component attributes in HoloCast, respectively, with different graph Laplacian matrix and hyperparameter as a function of wireless channel quality.
The key results from these figures are summarized as follows:
\begin{itemize}
\item The random-walk Laplacian matrix with the hyperparameter of standard deviation achieves the best performance in 3D coordinate attributes while the regular Laplacian matrix with the hyperparameter of variance yields the best quality in the color component attributes.
\item When we use the normalized and transition matrices as the graph Laplacian operator, the sender should use the standard deviation as the hyperparameter.  
\item Interestingly, HoloCast with the random-walk Laplacian matrix achieves better 3D coordinate reconstruction using the hyperparameter of the variance, while achieving high-quality color components reconstruction using the hyperparameter of the standard deviation.
\end{itemize}
How to optimize weight matrix and Laplacian matrix is still an open problem. We leave rigorous analysis as future work.

\subsection{Impacts of Different Point Clouds}
\label{sec:diff}
Previous sections use relatively small number of 3D points $N=2731$ to demonstrate the benefit of HoloCast.
In this section, we consider a larger number of 3D points as the test point cloud to show the scalability of the proposed HoloCast.  
Figs.~\ref{fig:snrvsmse_9_0}~(a) and (b) show the MSE of 3D coordinate and color component attributes in HoloCast, respectively, for the point cloud data of {\it pencil\_\_9\_0} ($N=6712$).
Compared with a small number of 3D points in Figs.~\ref{fig:snrvsmse}~(a) and (b), HoloCast achieves better reconstruction quality in both 3D coordinates and color components.
For example, HoloCast achieves $15.6$~dB and $2.4$~dB improvement on average across the channel SNRs between 0~dB and 30~dB, respectively.

\begin{figure}[t]
\centering
 \subfloat[Original]
 {\includegraphics[width=0.46\hsize]{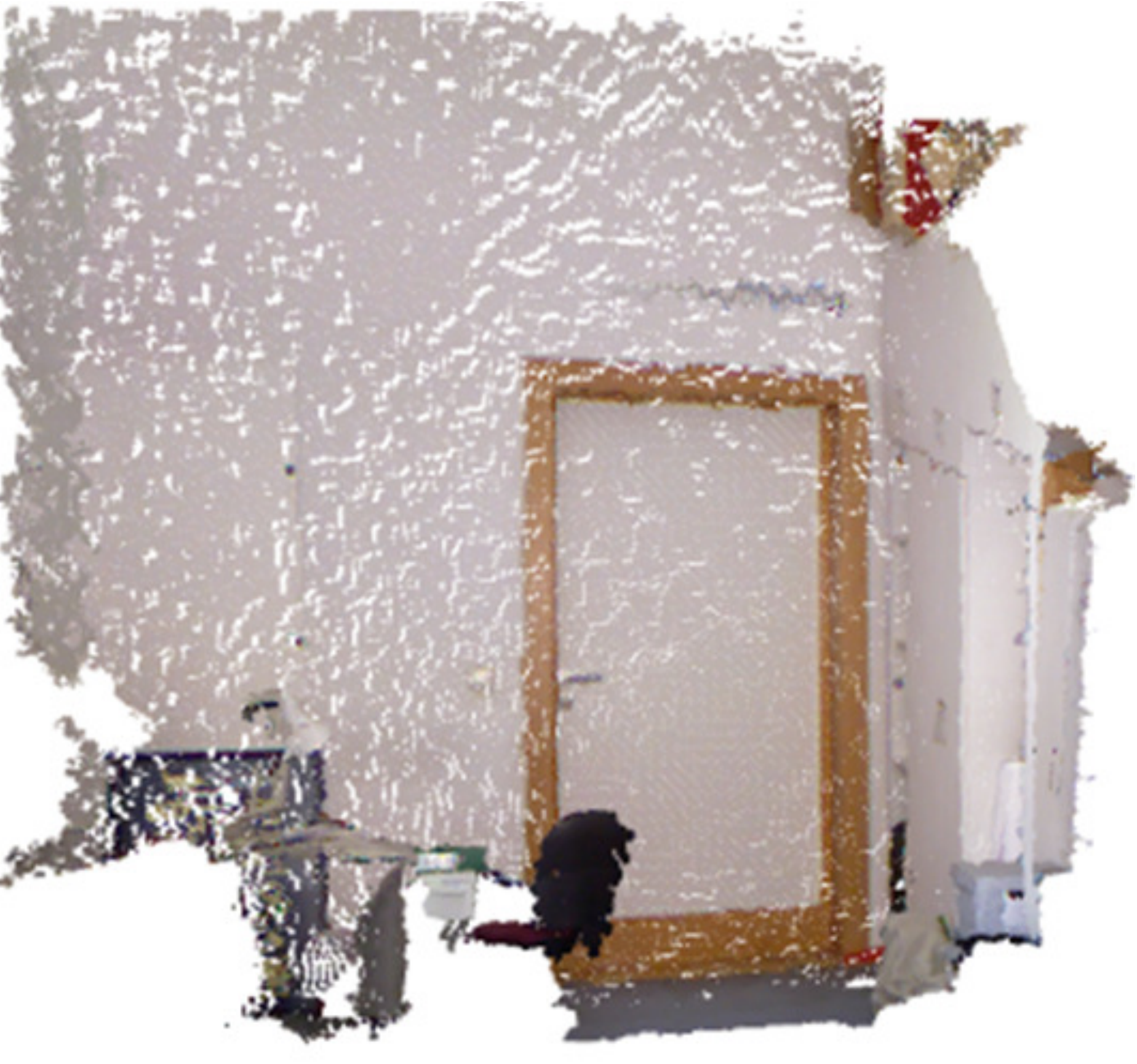} \label{fig:pantomime_interA}}
 \hfill
 \subfloat[Digital QPSK (SNR: 10dB) \newline $\mathsf{sMSE}_\mathsf{xyz}$: $-50.75$~dB \newline $\mathsf{sMSE}_\mathsf{yuv}$: $-12.93$~dB]
 {\includegraphics[width=0.48\hsize]{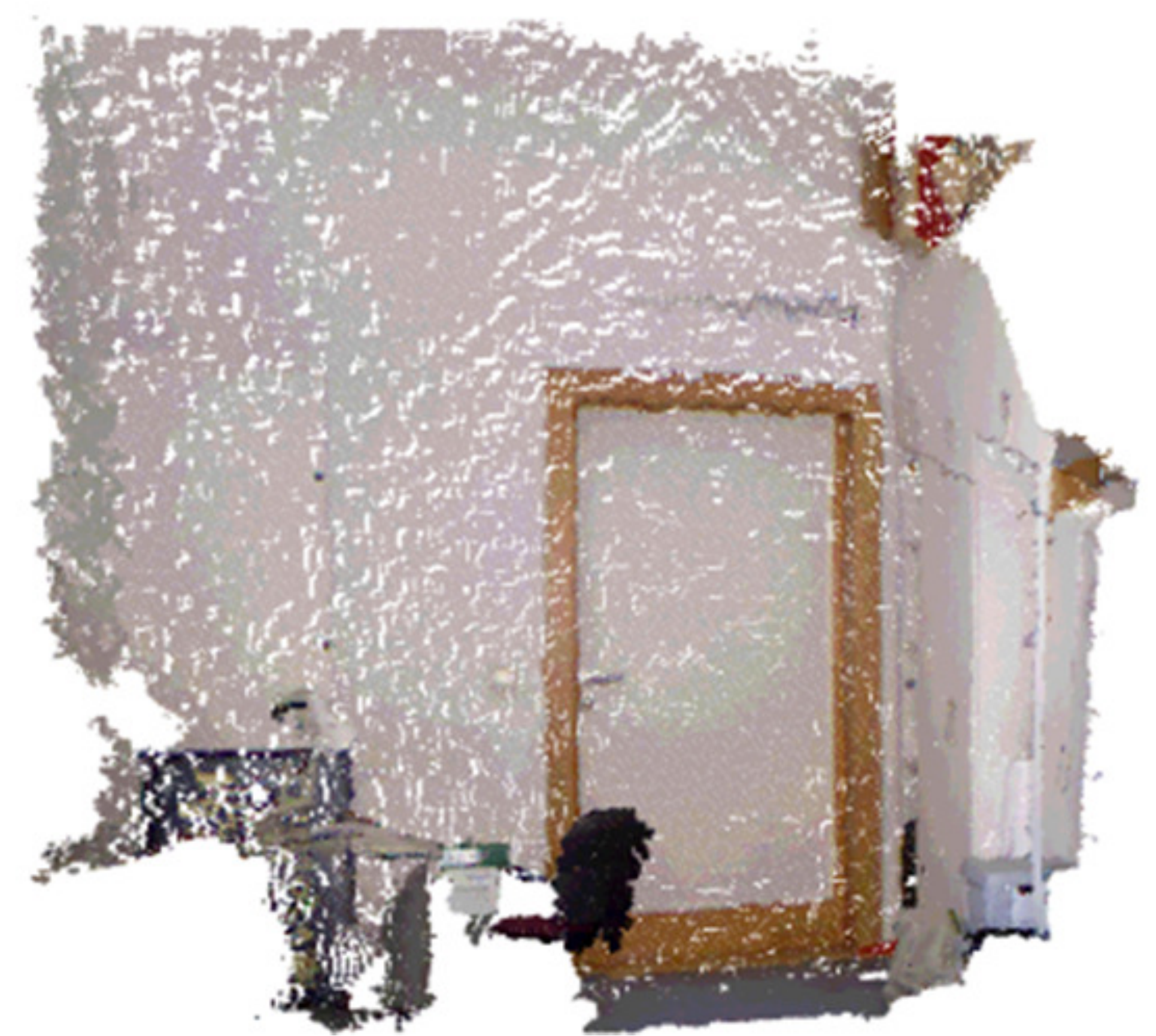} \label{fig:interB}}
 \\
 \subfloat[HoloCast (SNR: 10dB) \newline $\mathsf{sMSE}_\mathsf{xyz}$: $-40.47$~dB \newline $\mathsf{sMSE}_\mathsf{yuv}$: $-56.38$~dB]
 {\includegraphics[width=0.46\hsize]{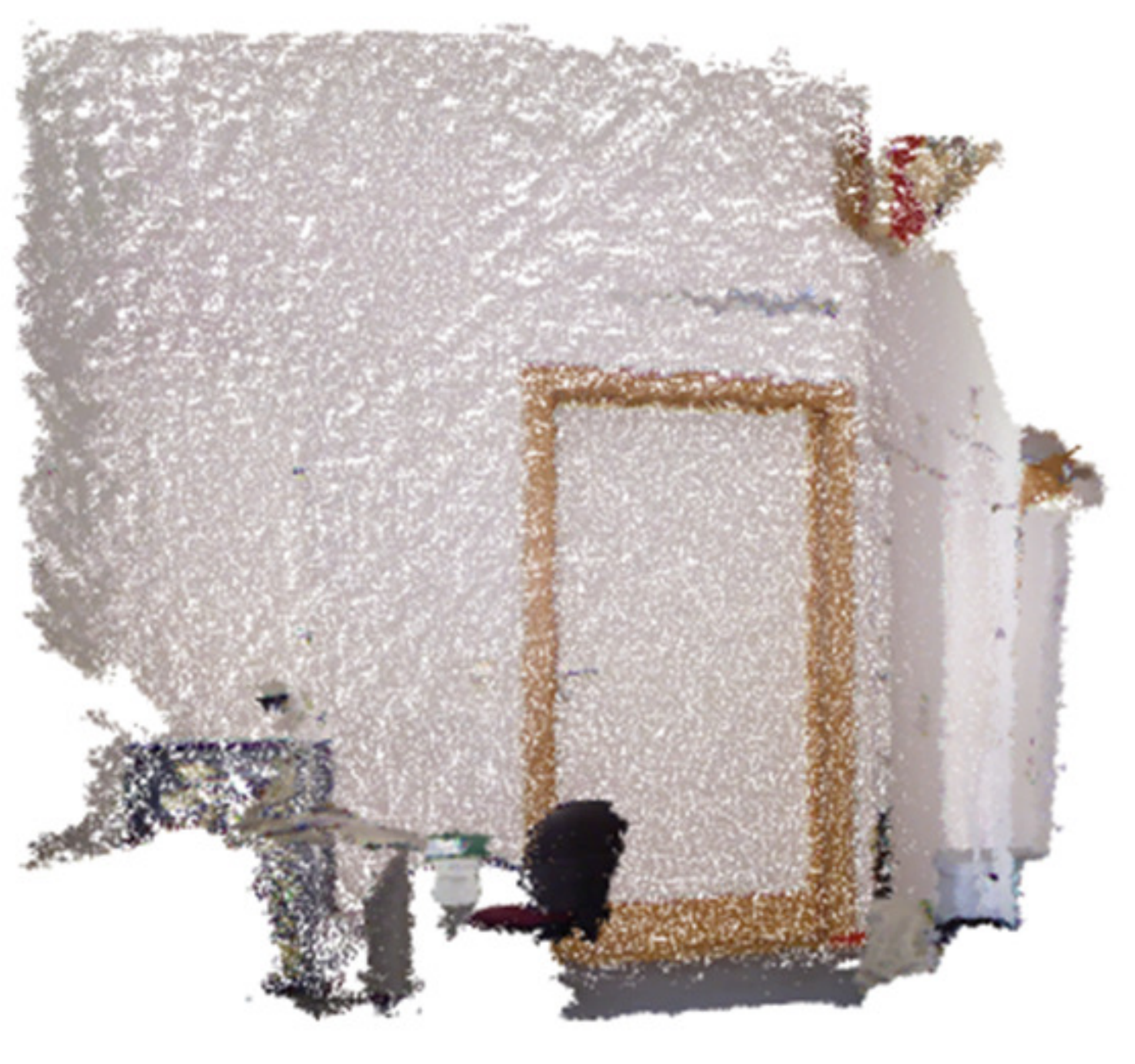} \label{fig:interD}}
 \hfill
 \subfloat[HoloCast (SNR: 20dB)  \newline $\mathsf{sMSE}_\mathsf{xyz}$: $-55.85$~dB \newline $\mathsf{sMSE}_\mathsf{yuv}$: $-72.01$~dB]
 {\includegraphics[width=0.46\hsize]{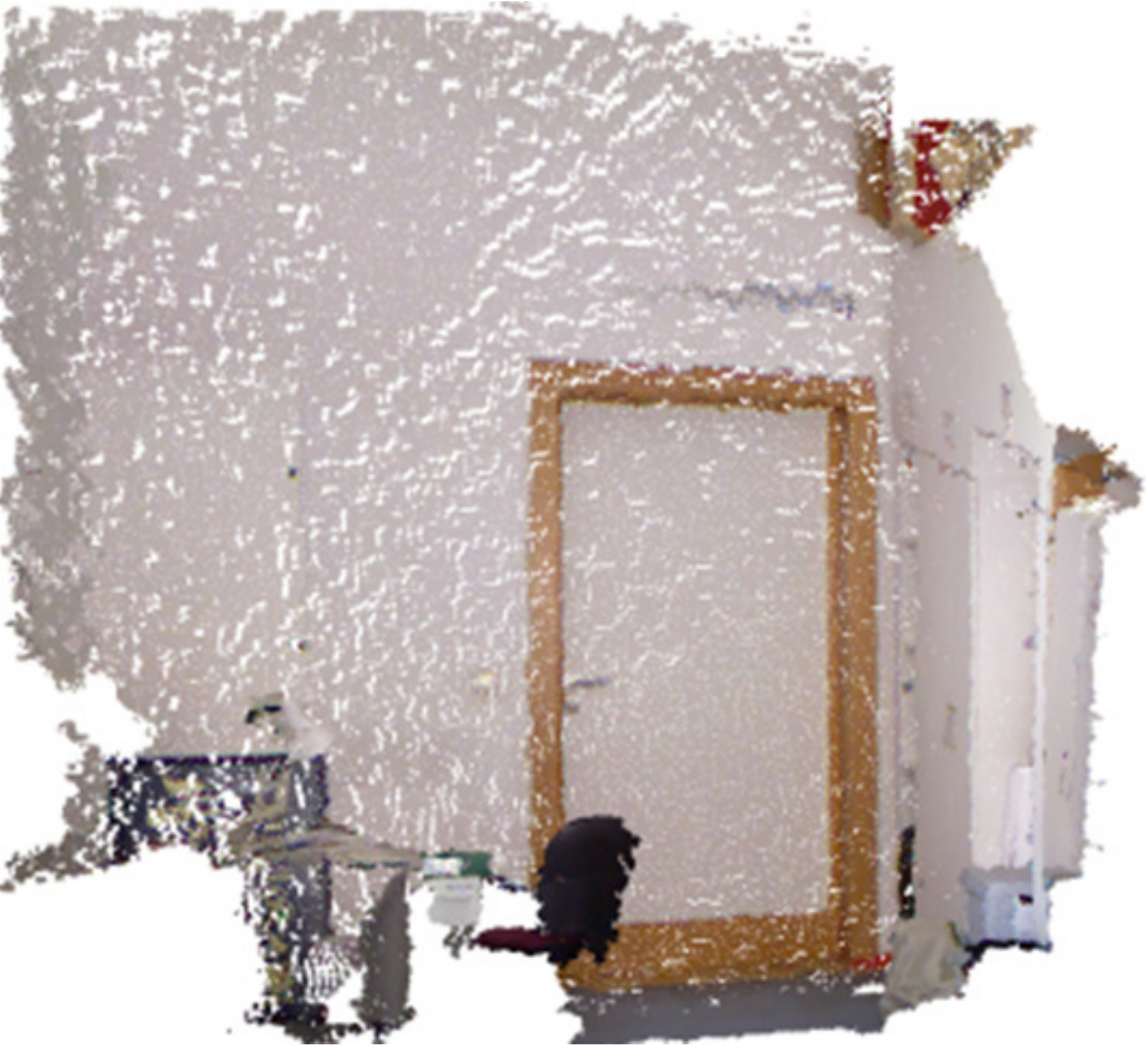} \label{fig:interE}}
 \caption{Snapshot of {\it office1} in digital-based and HoloCast schemes.}
 \label{fig:office1_frames}
\end{figure}

\subsection{Visual Quality}
\label{sec:visual}
Finally, Fig.~\ref{fig:office1_frames} compares the visual quality of HoloCast and digital-based schemes for the reference point cloud of {\it office1}.
We consider the digital-based scheme with QPSK modulation format at a channel SNR of $10$~dB, where the compressed bitstream can be successfully transmitted to the receiver over wireless channels. Here, HoloCast uses DCT for the decorrelation of the point cloud. 
The MSE of color attributes achieved by the digital-based scheme is $-12.93$~dB, whereas $-56.38$~dB and $-72.01$~dB are achieved by HoloCast at wireless channel SNRs of $10$~dB and $20$~dB, respectively. 
From the snapshots, we can observe that the digital-based scheme provides lower-quality point cloud (color degradation at the door). In contrast, HoloCast gracefully improves the reconstruction quality according to available wireless channel quality. Specifically, HoloCast can reproduce a clean 3D scene with details at a higher channel SNR of $20$~dB.

\section{Conclusion}
In this paper, we proposed HoloCast to realize graceful point cloud delivery over wireless links/networks.
In contrast to conventional 2D images, 3D point cloud data are not ordered and are non-uniformly distributed in space.
HoloCast regards the 3D points and color components as graph signals and directly transmits linear-transformed signals based on GFT.
Evaluation results with several point cloud data showed that HoloCast yields better reconstruction quality even at low wireless channel SNR regimes.
Feasibility study over practical experiments for various datasets with reduced amount of metadata will be conducted as future work.

\section*{Acknowledgment}
T. Fujihashi's work was partly supported by JSPS KAKENHI Grant Number 17K12672.

\bibliographystyle{IEEEtran}
\bibliography{./hybrid}

\end{document}